\documentclass[aps,prl,twocolumn,groupedaddress,dvips]{revtex4}
\usepackage[dvips]{graphicx}

\begin{document}
\preprint{}

\title{Nanometer-Scale Metallic Grains Connected with Atomic-Scale Conductors}
\author{A. Anaya, A. L. Korotkov, M. Bowman, J. Waddell, and D. Davidovic,} 
\email[]{dragomir.davidovic@physics.gatech.edu} 
\affiliation{Georgia Institute of Technology, Atlanta, GA 30332}
\date{\today}
\begin{abstract}
We describe a technique for connecting a nanometer-scale gold grain to leads by atomic-scale 
gold 
point contacts. These devices differ from previous metallic quantum dots in 
that 
the conducting 
channels are relatively well-transmitting. 
We investigate the dependence of the Coulomb blockade on contact resistance.  
The high-resistance devices display Coulomb blockade and the low-resistance 
devices display a zero-bias conductance dip, both in quantitative agreement 
with theory. 
We find that in the intermediate regime, where the sample resistance is 
close to $h/e^2$, 
the I-V curve displays a Coulomb staircase with symmetric 
contact capacitances. \end{abstract}
\pacs{73.23.-b,73.63.-b,73.22.-f } 
\maketitle

\section{introduction}
 
Fabrication of nanometer scale devices has recently become a subject of intensive study. 
In nanometer-scale metallic devices, 
discrete electronic energy levels can be resolved 
at low temperatures.
The effects of discrete electronic levels on quantum transport in small metallic samples
have 
been observed in atomic contacts and in nanometer scale grains.

Atomic contacts have been created using scanning probe microscopes~\cite{during} 
and mechanically controlled break junctions~\cite{moreland,muller}. 
In these techniques, a metallic contact is broken, for example by elongation, 
and one measures 
the conductance versus elongation. 
The combination of two effects, conductance quantization~\cite{wees,wharam} 
and sudden rearrangements of the atomic structure~\cite{landman,sutton}, 
cause the conductance to decrease in discrete steps while the contacts are 
stretched to the breaking point. The steps are on the order of the conductance quantum 
$G_0=2e^2/h$. 
 
In metallic grains, a single grain has been placed 
in weak tunneling contact with source and drain leads,~\cite{ralph,drago1} 
creating a single electron transistor (SET).~\cite{fulton} 
At dilution refrigerator temperature, discrete energy levels 
of the grain are measured from the I-V curve. The discrete energy spectra 
are analogous to those in artificial atoms.~\cite{kastner,ashoori,leo}

Study of both atomic scale contacts and grains
has led to major advances in understanding quantum transport in metals. 
For an extensive review of quantum properties of atomic-scale conductors 
and grains see Refs. ~\cite{ruitenbeek} and ~\cite{delft}, 
respectively.

In this paper, we describe a technique to
connect one gold grain to drain and source by 
atomic scale contacts. The samples that we generate 
demonstrate behaviors of both atomic contacts and metallic grains.
We show that the contact conductances can be changed in discrete steps 
of size comparable with $G_0$, showing that the conductance channels are 
well transmitting. 
This property differentiates our new devices from previously
studied grains.~\cite{delft} 
We investigate the Coulomb blockade of a grain as 
a function of the contact conductance. We identify a regime of intermediate 
Coulomb blockade, 
where the sample resistance is close to the resistance quantum. 

\begin{figure} 
\includegraphics[width=0.45\textwidth]{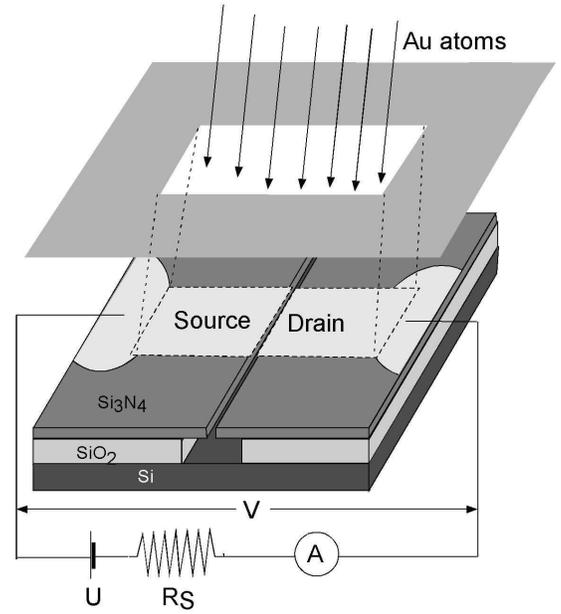} 
\caption{Metal deposition process for creating a point contact and/or a 
grain connected between the 
source and the drain. The gap between $Si_3N_4$ substrates is 70nm, the 
width of 
the undercut in $SiO_2$ is 120nm.\label{fig1}} 
\end{figure}

\section{Experimental setup}

The samples are created by combining metal deposition and electric field induced
surface migration of gold. 
Recently, an electromigration technique
has been implemented in fabrication of metallic electrodes with nanometer separation by Park et al.
~\cite{hongkun} In this technique a gold
nanowire had been made by electon-beam lithography, then broken by passing large current through it. The current flow had caused the electromigration of gold atoms, and the nanowire to break.
An electromigration technique was subsequently
applied at cryogenic temperatures,~\cite{jiwoong} which lead to the 
fabrication of atomic scale transistors.

Our technique is different from that of Park et al., in that a strong electric field 
is applied to create a connection between two separate  gold leads; it 
is based on electric field induced surface migration,~\cite{mayer} 
which is different from electromigration.

A $5N$ purity gold film 
is 
deposited on to two insulating $Si_3N_4$ substrates 
separated by a 70 nm slit, as shown in Fig. \ref{fig1}, in vacuum of 
$\sim 10^{-6}$ torr. The slit has a large undercut in $SiO_2$, 
which prevents electrical contact between the films. The exposed length of the 
slit is $0.1$ mm.

Electric current across the slit is monitored during the 
deposition. The applied voltage is $U$ and the 
voltage source impedance is $R_S$. 
When the thickness of the film is near $\sim 80 nm$, an electric 
contact between drain and source appears abruptly.

The contact between the drain and the source forms at a random location along the exposed length of the slit. 
The nature of the contact 
depends on $R_S$, $U$, the amount of gold deposited after the contact is detected, 
vapor pressure and temperature. The devices that can be obtained by this technique are
metallic point contacts, tunneling junctions, and grains.

\section{Metallic point contacts}
\label{metallic}

To create a metallic point contact, we set $R_S =0$ and $U\le 0.1$ V. Once the slightest conductance is detected, deposition is stopped.
With further gold deposition the conductance would increase rapidly
and display discrete steps in 
conductance versus time, of size $\sim G_0$.

In some samples, stable contacts with conductance
comparable with $G_0$ can be obtained by stopping the metal
deposition at the right time. These contacts 
are stable for minutes or longer.

Frequently, it is impossible to stop the deposition when the contact conductance is near $G_0$,
because the contact conductance jumps to a value of $\sim 100 G_0$. 
In addition, the contacts that have 
conductance of order $G_0$ often suddenly jump into this
$100 G_0$ state or they disconnect. 

\section{Planar Tunneling Barriers}
\label{tunneling}

\begin{figure} 
\includegraphics[width=0.45\textwidth]{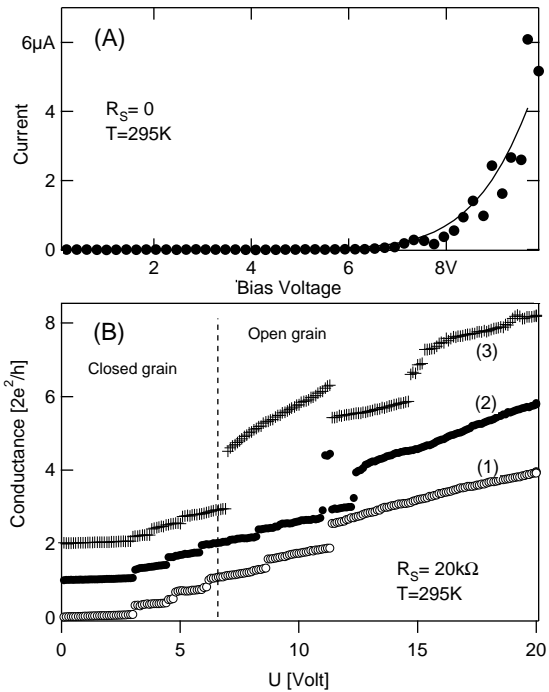} 
\caption{A. Circles: $I-V$ curve of a contact taken immediately after 
detecting 
the 
slightest current at 10 V bias. Line: Fit to an $I-V$ curve of a tunneling 
junctions. 
The fit parameters are $S=2\times 10^3nm^2$,$d=7.5\AA$, and $\Delta=6.4$ eV for 
the junction area $S$, 
gap size $d$, 
and the barrier height $\Delta$. 
B. Increasing the contact conductance by increasing $U$ in Fig.~\ref{fig1} 
with 
$R_S=20k\Omega$. 
Three curves are obtained in three samples. The curves are offset by $2e^2/h$ 
in vertical direction for clarity. These conductance traces are analogous to 
those found in break-junctions.~\cite{ruitenbeek}}
\label{fig2} 
\end{figure}

Intuition suggests that if two gold surfaces are in
mechanical contact, then there should also exist a good electrical contact. 
In this section, we show that mechanical contact does not necessarily
imply a good electrical contact.
It is possible to create a stable planar tunneling 
junction between two gold surfaces, with large tunneling resistance. 
This tunneling barrier is created if
the gold surface contains adsorbates.

Planar tunneling barriers between gold surfaces
have been first proposed by Hansen et al.~\cite{hansen,hansen1}
The authors observe that gold metallic contacts of conductance $\sim G_0$ 
grown in ultrahigh vacuum
are much less stable than
intentionally contaminated contacts with the same conductance. The enhanced 
stability is explained by a tunneling barrier composed of adsorbates,
which has a
mechanical contact area much larger 
than that for a metallic contact with the same conductance. 
The dimensions of these barriers are obtained by fitting the I-V curves. 
They vary from $\sim 12nm$ to $\sim 50nm$. The thickness of the barrier is of order several $\AA$.

We confirm this proposal in our experimental setup and clarify 
the conditions under which different types of contacts form.
Since the pressure during the metal deposition is 
$10^{-6}$ Torr, the surfaces contain adsorbates.
The nature of the contact that forms at this pressure
depends on the applied voltage.

If the applied voltage is weak ($U<0.1$ V), then the contacts are
metallic, as described in the previous section.
If the voltage is large, ($U>5$ V and $R_S=0$), the device contains a 
planar tunneling barrier and there is no direct 
metallic connection between drain and source. 

If the applied voltage is between $1$ V and $3$ V,
the contacts are mixed between the above two cases. They contain a large tunneling barrier and a small
metallic contact somewhere inside the barrier.
These contacts
have the advantage of having high stability and high channel transmittance. 
They are described in a separate section.

We discuss the high voltage regime first. 
We return to the deposition
process in Fig.~\ref{fig1} with $U=10$ V and $R_S=0$. 
When the $Au$ film thickness reaches $\sim 80$ nm, 
the current suddenly jumps to a nonzero value. 

The conductance $(I/V)$ 
strongly fluctuates in time. The average $I/V$ is much smaller than $G_0$. 
As soon as the contact is detected, we stop metal 
deposition and  
reduce the bias voltage close to zero. We minimize the time 
that the devices are exposed to $10$ V to
about one second.

While reducing the 
voltage, at a rate of one volt in ten millisecond, we measure the I-V curve. 
A typical I-V curve 
is shown in Fig.~\ref{fig2}-A. It is relatively well modeled by the I-V 
curve 
of a single tunneling junction~\cite{WOLF} with the barrier height 
comparable to 
the gold work function $W_{Au}=5.1$ eV. The diameter of the barrier is $\sim 10$ nm,
consistent with ref.~\cite{hansen,hansen1}.

The barrier height
is slightly  
larger than $W_{Au}$. This is explained by significant contamination of the junctions with
adsorbates, such as $H_2O$.~\footnote{
In a separate publication, we examine the effect of water molecules 
on the I-V curve in Fig.~\ref{fig2}-A in detail. We find that water is an excellent 
insulator:
when water vapor pressure is increased, the barrier height 
increases, increasing the sample resistance
by orders of magnitude.}

We image the contacts by the scanning electron microscope (SEM).
Transferring the sample 
from the deposition chamber to another instrument 
almost certainly leads to a sudden increase in resistance 
to an undetectably high value, despite being grounded at both ends during
transfer. 
In most samples, the electrical contact can be reestablished 
by applying a one second long voltage pulse at 10 V (or sometimes higher)
in high vacuum. The sample returns to the high resistance
tunneling barrier.

\begin{figure} 
\includegraphics[width=0.45\textwidth]{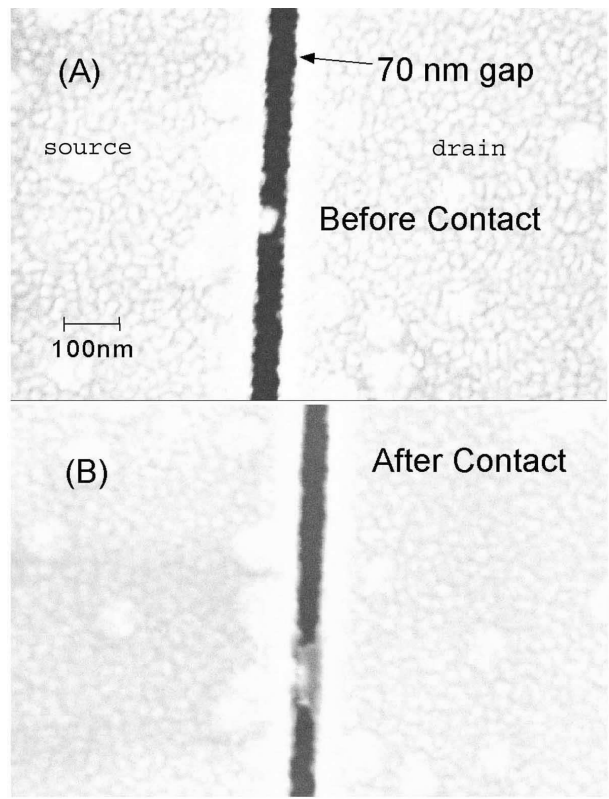} 
\caption{A. One gold grain attached to one lead. 
B. The grain after applying the 10 volt pulse.\label{fig3}} 
\end{figure}

Figure~\ref{fig3}-A shows one gold grain in
mechanical contact with the source. The sample disconnected
spontaneously while transferring to the SEM as described above. 
We apply the voltage pulse, while imaging 
and measuring the current simultaneously. 
The current starts to flow during the pulse duration. 
After the pulse, 
the sample resistance at low bias is $\sim 1G\Omega$ or higher.

Figure ~\ref{fig3} B shows the device after the voltage pulse. 
We verify through microscopy that no 
additional connections are formed outside the slit in Figure~\ref{fig3}. 
The mechanical contact area 
is of order $10$ nm, confirming the proposal by Hansen et al.

Under the influence of voltage, the grain pulled a Au protrusion from the 
right lead, establishing contact. 
This protrusion forms because of a process known as electric field induced 
surface migration, see Ref.~\cite{mayer} and references therein. 

It is understood that
the protrusion grows because of a force acting on
surface gold atoms. 
When the electric field 
exceeds a certain threshold, it leads to migration of surface gold atoms toward the region of the strongest electric field, creating a protrusion.
The protrusion increases the electric field gradient,
which in turn accelerates its growth. When the protrusion reaches the other side,
a planar junction is formed. This junction is stable after the voltage is reduced.


\section{Formation of grains  in electric breakdown}

Consider a metallic point contact with resistance less than $10\Omega$. 
We apply a voltage pulse of 10 V in vacuum. 
The contact breaks down, and after the pulse the resistance is immeasurably large. 

We observe the break-down by the SEM.
Fig.~\ref{fig4}-A shows 
an area near the contact after the break down. 
Evidently, a relatively wide region around the point contact 
melts under the influence of the voltage pulse.
The molten gold retracts away from the slit, presumably in order to reduce surface 
energy.

In addition, a large number of electrically isolated grains is left 
behind 
on the substrate surface. This is shown in fig.~\ref{fig4}-B. 
The image indicates a broad distribution of grain diameters. 
The particles have nearly circular shape. This suggests that the 
particles have been molten (in the liquid state, minimizing the 
surface energy in contact with the substrate leads to circular grains).

\section{Planar Tunneling Barriers - Revisited}


In Fig.~\ref{fig3}-B, there is a grain between drain and source.
The majority of devices fabricated with a 10 V pulse, as in fig.~\ref{fig3}-B, 
have I-V curves that exhibit the 
Coulomb staircase at low temperatures. Since the 
staircase is conditional on $R_L,R_R>h/e^2$, where $R_L$ and $R_R$ are the contact resistances between the grain and the leads, this shows that
there are at least two contacts in series between drain and source. 
In later sections, when we discuss Coulomb Blockade,
we infer from the I-V curves that there is only one grain
between the drain and the source.
Figure~\ref{fig3}-B supports this conclusion. 

\begin{figure} 
\includegraphics[width=0.45\textwidth]{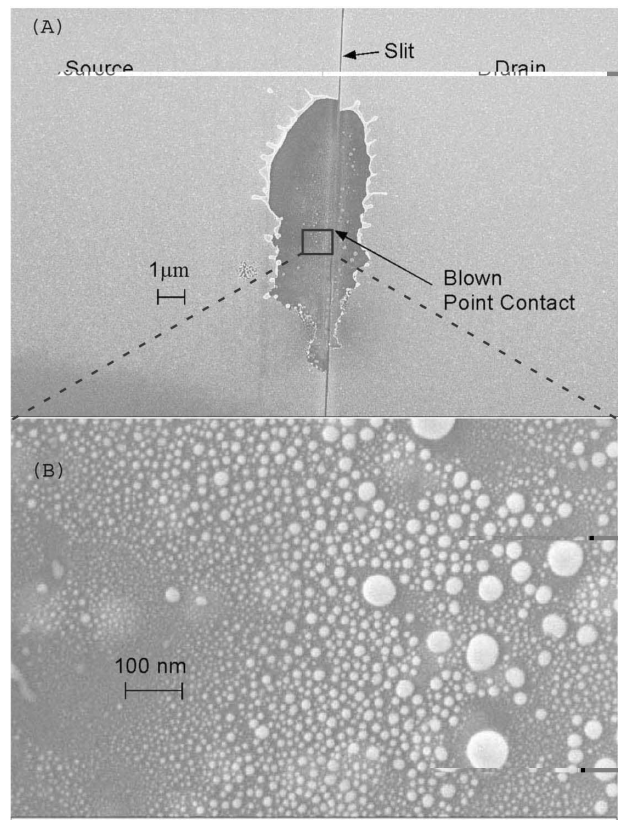} 
\caption{A. Scanning Electron Microscope image of a metallic point contact 
after 
the electric breakdown by the voltage pulse, as described in the text B. A 
zoomed-in image of the blown region on the left side of the slit. 
An array of circular gold grains is left on the substrate.\label{fig4}} 
\end{figure}

We propose the following scenario for grain formation.
We suggest that the very first contact at 10 V bias voltage
is metallic. It lasts only for a very short time, because it breaks down by melting.
During the break down, one or several grains are created analogous to Fig.~\ref{fig4}.

\section{Mixed Contacts}
 
After
creating a device with a 10 V pulse, as in fig.~\ref{fig3}-B,
we change the circuit in Fig.~\ref{fig1}, so that $R_S=20k\Omega$
and
slowly increase $U$ from zero, and measure the conductance ($I/V$). 
Note that $U$ is the total voltage across the sample in series with $R_S$. 
Figure~\ref{fig2}-B 
shows the resulting conductance traces in 3 different samples. 

$I/V$ jumps to a significant fraction of $2e^2/h$ at roughly $2.8$ V. After the jump, the voltage 
across the sample $V$ is reduced below $U$ by voltage division with $R_S$, and 
the sample voltage becomes close to $1.7$ V. 
Further increase in $U$ results in two effects: there are additional discontinuities 
in conductance, of amplitude $0.2$ to $2e\,^2/h$, and there is a smooth increase 
in conductance versus $U$. The increase in conductance is maintained if the voltage $U$ is removed.

The discontinuities in conductance of size $\sim e^2/h$ 
suggest that discrete atomic reconfigurations take place, 
analogous to those in atomic conductors.~\cite{ruitenbeek} 
The steps 
demonstrate that the contacts contain well transmitting channels. 
In  Fig.~\ref{fig3}-B, the atomic scale contacts form somewhere inside the 
tunneling barrier between the grain and the leads. 


The smooth 
conductance increase suggests that the
tunneling contribution is increasing.
We confirm through microscopy that the area of the planar barrier is increasing while the contact conductance 
is increasing smoothly.
 
As described before, in a strong nonuniform 
electric field, at room temperature, surface gold atoms 
move toward the strongest electric field.~\cite{mayer} 
We propose that when the electric field strength increases beyond a certain point,
one atom pushes through the barrier, creating a metallic contact. 

Once the first atom is connected, 
further increase in the electric field adds 
more gold atoms into the contact, increasing the number of channels and 
the conductance in steps of size $\sim e^2/h$.

These discrete atomic additions occur in both barriers. 
In fact, since the 
voltage drop is stronger at the higher resistance junction, and voltage drop drives the
conductance increase, it follows that
the increase in conductance tends to even out any imbalance between the two 
resistances. 
This balancing of the electric field leads to the desired property that 
the Coulomb blockade at low temperature 
vanishes in a relatively narrow range of 
sample resistance. 

\section{Coulomb Blockade}
To observe Coulomb Blockade, a sample whose resistance at 
room temperature is greater than or equal to 20 k$\Omega$ must be cooled 
to milliKelvin temperatures.  However, the evaporator in which each 
sample is prepared cannot be cooled.  So each sample must be moved to 
a dilution refrigerator where it can be cooled to as low as 10mK.

	The samples are very sensitive to electrostatic shock and 
so a careful grounding procedure must be followed to safely transfer 
each sample from the chamber to the refrigerator.  From outside the 
chamber, before venting to atmosphere, the voltage on the sample is 
reduced to zero, and both leads are grounded.  After venting, a 
mobile grounding strap grounds the removable deposition stage, 
and the sample leads are grounded to the stage, where now the 
voltage leads to the outside circuitry may be safely removed.  
The stage may now move to an ESD-safe workstation (table) and 
the sample dismounted.

	To dismount the sample from the stage, a small piece 
of indium wire is used to shunt the contact pads on the chip.  
Leads running from the stage to chip may now be disconnected.  
The chip is free to move to a stage more suitable for the dilution refrigerator.  
A small chip carrier, designed for use in the tailpiece of the refrigerator will house the sample.  
Leads from the carrier to the chip are connected by indium pressing to the contact pads.  The leads, 
from the top of the refrigerator to a connector at the tailpiece, are grounded, and the carrier is 
finally attached to the connector.  The indium shunt on the chip is removed.  The sample is 
now available for cool down.  Only through this laborious grounding procedure can we move 
samples and have them survive.

We have recently added the gate to our samples. 
To this end, the mask is placed back over the chip,
but rotated by $90$ degrees. Then the sample is returned to the deposition chamber and reconnected to the outside electronics. If the resistance of the sample is infinite,
the sample is reconnected in high vacum, as in Fig.~\ref{fig3}. Then, we deposit a 
layer of $Al_2O_3$, by reactive evaporation of aluminum.~\cite{drago1} 
Aluminum is deposited at 
angles of $\pm 5\deg$ relative to the substrate normal. For each angle, the deposited thickness is 15nm. Then, we deposit a layer of aluminum metal, 40 nm thick, 
in the direction parallel with the normal.

The refridgerator is closed and
pumped down to $\sim 10^{-5}$ Torr. Then, 
the sample resistance is modified (if needed), using the techniques described in previous sections. Finally, the sample
is cooled to $0.015$ K and the I-V curves are measured.

Standard cryogenic filtering techniques lower 
the electron temperature. Differential conductance is 
obtained using a lock-in technique. 
The applied voltage is a sum of a constant dc-voltage $V$ 
and 
a weak ac-voltage $v(t)=v_{AC}\sin(2\pi ft)$, where $v_{AC}=1\mu$ V
and 
$f=500$ Hz. Current $I$ is measured with an 
Ithaco current preamplifier. To obtain the differential conductance, the 
output of the amplifier 
is connected to the input of a lock-in amplifier. 

\begin{figure} 
\includegraphics[width=0.45\textwidth]{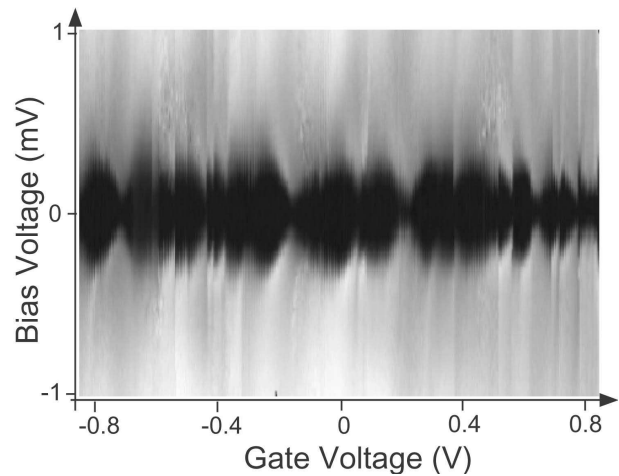} 
\caption{Gate voltage dependence of the Coulomb blockade, in sample 3. Darker = smaller conductance. The diamonds represent different charged states of the grain, 
analogous to the gate voltage 
dependences in quantum dots.~\cite{leo}}\label{fig5} 
\end{figure}

The effects of the gate voltage on differential conductance are shown
in fig~\ref{fig5}, at $T=0.015K$ in sample 3. The differential
conductance near zero bias voltage is strongly suppressed.
If the bias voltage aproaches a certain threshold from below, the differential conductance rapidly increases by several orders of magnitude.

The voltage threshold displays 
quasiperiodic modulation with the gate voltage. 
By applying the gate voltage, 
the voltage threshold can be reduced 
to zero. If the gate voltage is chosen so that the voltage threshold is at maximum, 
then the conductance at zero bias voltage is four orders of magnitude smaller 
than the conductance at large bias voltage.
 
This proves unambiguously that the sample consists of a metallic grain 
connected to drain and source, rather than a single junction. The diamonds 
in Fig.~\ref{fig5} are symmetric around
zero bias voltage, 
showing that $C_L\approx 
C_R$, where $C_{L,R}$ are the effective capacitances
determined from the fit to the Orthodox theory 
of single-charge tunneling on a single metallic 
island.~\cite{likharev} 
The image is chopped, because of the random switches in the 
background charge distribution (note that the gate voltage range is 
rather large compared with the bias voltage).

\begin{figure} 
\includegraphics[width=0.45\textwidth]{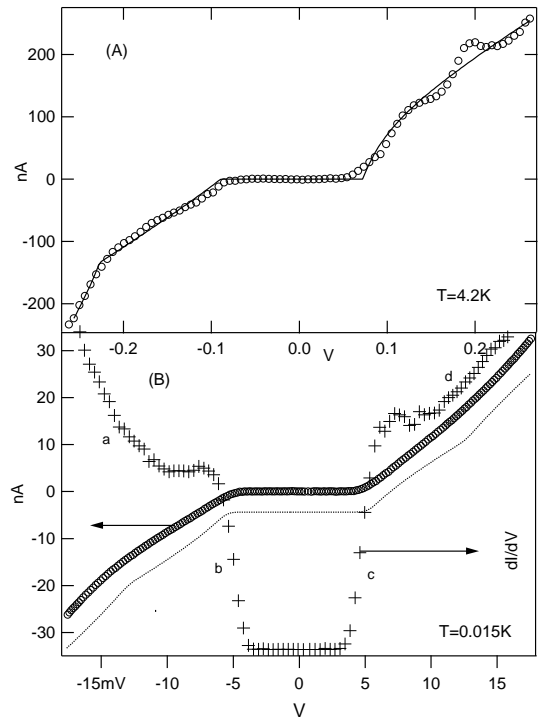} 
\caption{A. Circles: I-V curve of sample 1. Line: fit to the 
Coulomb staircase on single grain obtained from the Orthodox theory. 
The charging energy $E_C=82meV$. 
B. Circles: I-V curve of sample 2. Crosses: differential conductance versus 
voltage. Line: fit to the Orthodox theory. The line is offset in vertical direction for clarity.
The best fit parameters 
determine effective capacitances, not the geometric capacitances, as described 
in the text.}\label{fig6} 
\end{figure}

Figure~\ref{fig6}-A shows the 
I-V curve of sample 1. This I-V curve is typical for our high resistance 
samples. 
We fit the I-V curve 
using the Orthodox theory,~\cite{likharev} 
and show the corresponding fit by the dashed line. Good fits 
are generally found in approximately $50\%$ of the samples, suggesting that 
the samples typically consist of a single metallic grain in 
weak electric contact with the drain and the source. 

The charging energy $82$ meV extracted from the fitting
is relatively large. As a result, the bias voltage 
can induce jumps in the background charge distribution (also known as 
$Q_0$-shifts in the Orthodox theory). 
The I-V curve of sample 1 displays a noticeable deviation from the fit at 
positive bias voltage, see Fig.~\ref{fig6}-A. These deviations are not noise. 
They are hysteretic - they change when sweeping the voltage up and down. At 
negative voltage bias, the deviations from the fit are weak and we do not 
observe any hysteresis when repeating the voltage sweep. This behavior is 
thus far consistent with single charge tunneling.

Among samples, charging energy fluctuates strongly. 
These fluctuations are explained by the broad distribution of particle 
diameters possible in such systems, Fig.~\ref{fig4}. 
In addition, there is another source of fluctuations in charging energy, 
caused by the distribution of contact conductances. This effect 
will be discussed in the next section.

\begin{table} 
\caption{
$R_L+R_R$, $R_L/R_R$ , $C_L+C_R$, $C_L/C_R$: junction resistances and 
capacitances determined from the Coulomb staircase, in samples 1,2, and 
3, and from the 
theory of strong tunneling in an SET, in sample 4. In samples 1-3, the capacitances
are effective, and in sample 4, the capacitance is bare. $E_C$ is the effective charging energy.} 
\label{tab1} 
\begin{tabular}{|c|c|c|c|c|c|} 
\hline 
&$R_L+R_S [k\Omega]$&$R_L/R_S$&$C_L+C_S [aF]$&$C_L/C_S$& $E_C [meV]$\\ 
\hline 
1& 550 & 2.9 & 0.98 & 1.2 & 82\\ 
\hline 
2 & 344 & 1 & 17.4 & 1 & 4.6\\ 
\hline 
3 & 56 & 1 & 410 & 1 & 0.2\\ 
\hline 
4 & 14 & 7.1 & 5.52 & - & $\ll$ 0.1\\ 
\hline 
\end{tabular} 
\end{table}

\section{Intermediate Coulomb Blockade}

As $R_L+R_R$ decreases, 
the effective charging energy rapidly approaches zero, as shown in table 1.
If the gap in the $I-V$ curve is 
well resolved at $T=0.015K$, the grains are referred to as closed. The 
range of 
conductances of closed grains is indicated in Fig.~\ref{fig2}-B. 
In table 1, samples 1-3 display a clear Coulomb blockade 
at $T=0.015K$. The parameters in these samples are 
obtained from the fits to the orthodox theory.

Figure~\ref{fig6}-B shows $I-V$ and $dI/dV-V$ curves in sample 2. 
This sample displays a sharp Coulomb blockade, similar to that in sample 1. 
The I-V curve has no hysteresis, which is not surprising, since the charging 
energy is only 
$4.6$ meV and the bias voltage is too weak to reconfigure the background charge 
distribution. 
The Orthodox theory is in good agreement 
with the linear segments of the I-V curve. We can identify the same threshold 
voltages at both signs of the bias voltage: (a) corresponds to (d) and 
(b) corresponds to (c). This confirms there is only one grain in series 
with drain and source. 

The rounding of the I-V curve 
is larger than expected from the Orthodox theory. A good fit is obtained with identical junction parameters 
$R_L = R_R$ and $C_L=C_R$. 
The first identity, $R_L=R_R$, is not as 
important
as the second identity; good fits can be obtained even if 
$R_L/R_R\neq 1$. 
$C_L=C_R$ follows 
directly 
from the I-V curve, from the fact that the threshold voltages (b) and (c) have 
the same magnitude. 
We have noticed this symmetry in contact capacitance in a relatively large 
number of samples - roughly 
$50\%$ of samples with charging energy less than $5$ meV. 
In samples with $E_C<1$ meV, this symmetry is 
found in nearly every sample studied. 

We do not have an explanation for this symmetry at low temperature and low 
bias 
voltage. Experimentally, no special care has been taken to 
make the junctions symmetric. In fact, the parameters are expected to be 
significantly asymmetric. 
We note that the symmetry in effective capacitance
is broken in a strong magnetic field. The discussion of the magnetic field dependence is beyond the scope of this paper.

Theoretically, one expects that 
if $G_L+G_R>G_0$, where $G_{L,R}=1/R_{L,R}$,
the system parameters become renormalized 
near 
zero temperature and zero bias voltage. The effective charging energy 
of the grain decays exponentially with contact conductance, 
\begin{equation}
E_C^{eff}=E_Ce^{-\alpha (G_L+G_R)/G_0},
\label{eq1}
\end{equation} 
where $\alpha$ is a constant of 
order 1, dependent on the nature of the contacts ($\alpha = 1$ in tunneling junction and $\alpha =\pi^2/4$ in a diffusive metallic contact), 
$E_C$ is the bare charging energy.~\cite{zaikin,nazarov} This exponential
dependence on $G_L$ and $G_R$ 
also explains strong fluctuations in the charging energy 
among samples. 

The suppression of Coulomb blockade have been studied in larger metallic SETs.~\cite{joyez,chouvaev}
The connection between the effective charging energy and the bare 
sample parameters have been established. ~\cite{joyez} 
Our samples are different from metallic SETs in that
the contacts are metallic, and the grain diameter is small - small enough so that
an electron can enter and exit the grain without loosing its phase coherence.~\footnote{The traversal time is roughly $((R_L+R_R)/R_Q)h/\delta$, where $\delta$ is the level spacing and
$R_Q=h/e^2$. For a grain of diameter 20nm, the traversal time is easily shorter than 
the dephasing time in pure gold at low temperatures.}  

No calculation of the $I-V$ curve in this regime is available. We still
fit the $I-V$ curve 
using the Orthodox theory, with the notion that the best fit parameters 
represent 
effective (renormalized) parameters. In this sense, $E_C^{eff}=e^2/(2C^{eff})$. 
The central result of this paper, beyond describing the new technique, is that
$C_L^{eff}=C_R^{eff}$, when $E_C^{eff}\ll E_C$ and when the applied magnetic field is weak.
In table 1, the sample parameters are the effective parameters
as defined here. No connection between effective and bare
parameters is established in this paper. 
This connection is the subject of present research in our laboratory.

The observation $C_L^{eff}=C_R^{eff}$ 
has important implications for conversion from 
bias voltage into grain energy. For example, if the density of 
states of 
the grain varies as a function of energy, then this variation can be 
measured from the I-V curve. 
The capacitive division prefactor, which converts from bias voltage to energy, 
is 2.

In instances where the 
theoretical calculations can be 
solved exactly, 
the theory predicts that any asymmetry in the junction's 
real parameters is only enhanced by the renormalization 
at low temperatures and low bias voltage.~\cite{furusaki} The theoretical 
model, 
however, 
is valid only if there is one electronic mode per contact, which may not apply 
to our samples. Our data demonstrate the {\it opposite} behavior from that 
predicted theoretically: in a weak magnetic field,
the effective capacitances are symmetrized.


\section{Weak Coulomb Blockade} 
\begin{figure} 
\includegraphics[width=0.45\textwidth]{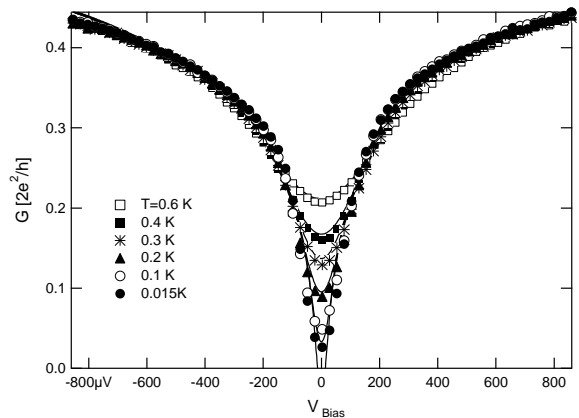} 
\caption{Symbols: Differential conductance versus bias voltage in sample 3 
as a function of temperature. Lines: Fit to the theory of strong tunneling 
in a 
metallic grain.}\label{fig7} 
\end{figure}

We have observed that when the sample resistance at room temperature is less 
than roughly 
$10k\Omega$, then the conductance at $T=0.015K$ near zero bias voltage 
remains on the same order of magnitude as the conductance at large bias 
voltage (asymptotic conductance/resistance). 
In Fig.~\ref{fig2}-B, the grains with this property are referred to as open.
Fig.~\ref{fig7} shows $dI/dV$ versus $V$ 
in a sample 
with room temperature resistance $\approx 14k\Omega$. We observe that 
the conductance at $V=0$ and $T=0.015K$ 
remains within an order of magnitude from the room temperature conductance. 
In samples with asymptotic resistance $10k\Omega<R_L+R_R<20k\Omega$,
the depth and the width of the conductance  dip fluctuates  
strongly among samples (in two out of forty samples studied thus far, 
the conductance dip is significantly less pronounced than that in Fig.~\ref{fig7}).

As the temperature increases, the zero-bias dip broadens. While at room 
temperature, 
the zero bias conductance dip is no longer resolved; the I-V becomes 
linear, with the conductance equal to the $1/(R_L+R_R)$. Alternatively, this 
asymptotic
conductance is obtained 
at low temperatures, by applying a large enough bias voltage 
so that the I-V curve approaches the linear form.

To interpret this result we first try to
fit the data in Fig.~\ref{fig7} to the dynamic Coulomb 
blockade model of 
a single tunneling junction.~\cite{devoret,girvin} Yeyati et al. 
have extended this model to include a
metallic contact with one channel,~\cite{yeyati}
showing that the temperature dependence of a metallic contact does not
differ from that of a tunneling junction. 
Fitting to the dynamic Coulomb blockade model
does not
reproduce the temperature dependence in Fig.~\ref{fig7}.

We fit the data in Fig.~\ref{fig7} to the theory of a mesoscopic metallic grain 
in the strong tunneling limit.~\cite{golubev} This theory has already been used 
in connection with measurements of strong tunneling SETs by Chouvaev et al.~\cite{chouvaev}
Strictly speaking, the model in Ref.~\cite{golubev} is not applicable to our devices because 
our contacts are metallic. For example, the parameter $\alpha$ in Eq.~\ref{eq1} 
differs between
metallic contacts and tunneling contacts by a factor of $\pi^2/4=2.46$. This implies
that the Coulomb blockade is much more strongly suppressed with metallic contacts than with tunneling contacts.~\cite{nazarov}.
In the absence of a theory of the I-V curve for a grain 
with metallic contacts, we use the theory
of tunneling contacts.~\cite{golubev}
The theory is valid only if 
$eV \gg E_C^{eff}$ or $k_BT\gg E_C^{eff}$.

The I-V curve is 
\begin{eqnarray} 
I(V)&=&\frac{V}{R_L+R_R}-\frac{e}{\pi\hbar}\frac{R_LR_R}{(R_L+R_R)^2}\,{\rm 
Im}\sum_{r=L,R} 
\nonumber\\ && 
\left(\frac{\hbar}{R_0C}+ieV_r\right)\Psi\left(1+\frac{\hbar}{2\pi 
k_BTR_0C}+i\frac{eV_r}{2\pi k_BT}\right)\nonumber\\ && 
-ieV_r\Psi\left(1+i\frac{eV_r}{2\pi k_BT}\right) 
\label{eq2}
\end{eqnarray}
Here $R_0=\frac{R_LR_R}{R_L+R_R},$ $C=C_L+C_R,$ $V_L=\frac{R_L}{R_L+R_R}V,$ 
$V_R=\frac{R_R}{R_L+R_R}V$ and $\Psi (x)$ is the digamma function. Note that there is a misprint in Ref.~\cite{golubev}. The corrected expression is given in Ref.~\cite{chouvaev}.

There are three fitting parameters $R_L,R_R$ and $C$. We set $R_L+R_R$ 
to be the same as the sample resistance at high temperature/voltage, 
and vary $C$ and $R_L/R_S$ to obtain the best fit. We fit the family of 
$dI/dV$ 
vs $V$ curves 
measured at different temperatures. The best fit is shown by lines in 
Fig.~\ref{fig7}, 
and the corresponding sample parameters are given in Table 1, for sample 3. 
The fit reproduces our data quite well in the entire temperature range. 

The I-V in Eq.~\ref{eq2} does not depend on 
on the capacitance ratio $C_L/C_R$. It is reasonable to assume that 
if the conductance 
at $V=0$ and $T=0.015K$ is several times smaller than $1/(R_L+R_R)$, then the 
voltage division across the grain is closer to capacitive than resistive. 
Then, if Eq.~\ref{eq2} holds,
it follows that the effective capacitive division is independent on
the bare capacitance ratio $C_L/C_R$. This may partially explain why the
measured capacitances in the intermediate Coulomb blockade regime are 
identical.

\section{Conclusion}
In conclusion, we present a new technique for fabrication of 
metallic grains in contact with leads. 
These grains are connected by two metallic atomic-scale 
point contacts, with well transmitted conduction channels. 
We show that when the sample resistance at room temperature is above $h/e^2$, 
then the $I-V$ curve at low temperatures displays the Coulomb blockade. 
When the 
sample resistance approaches $h/e^2$, the blockade is weakened. 
We discover an intermediate regime in which the I-V curve is described 
by the Coulomb staircase with symmetric 
junction capacitances. As the sample resistance is reduced further, the 
Coulomb blockade is 
completely washed out, and only a weak zero bias conductance dip is observed. 
This regime is 
well described by the theory of metallic grains with strong tunneling.
\begin{acknowledgments} 
We thank M. Pustilnik, L. P. Kouwenhoven, D. Golubev for useful 
discussions. We thank the 
Georgia-Tech electron microscopy facility where electron transport experiments 
in combination with sample imaging have been performed. This work was 
performed 
in part at the Cornell Nanofabrication Facility, (a member of the National 
Nanofabrication Users Network), which is supported by the NSF, under grant 
ECS-9731293, 
Cornell University and Industrial affiliates. 
This research is supported by the David and Lucile Packard Foundation 
grant 2000-13874 and the NSF grant DMR-0102960.
\end{acknowledgments}

\bibliography{zba} 

\end{document}